\documentclass{article}

\usepackage{arxiv}

\usepackage[hidelinks]{hyperref}       
\usepackage{nicefrac}       
\usepackage{doi}

\usepackage{amsmath,amsfonts}
\usepackage{graphicx}
\usepackage{textcomp}
\usepackage{xcolor}
\usepackage{booktabs}
\usepackage{listings}
\usepackage{multirow}
\usepackage{caption}
\usepackage{subcaption}
\usepackage{framed}

\usepackage{soul}
\usepackage[utf8]{inputenc}
\usepackage[T1]{fontenc}
\usepackage{microtype}
\usepackage{rotating}
\usepackage{paralist}
\usepackage{tabularx}
\usepackage{multicol}
\usepackage{pbox}
\usepackage{makecell}
\usepackage{enumitem}	
\usepackage{colortbl}
\usepackage{pifont}
\usepackage{xspace}
\usepackage{url}
\usepackage{tikz}
\usepackage{fontawesome}
\usepackage{lscape}
\usepackage{color}
\usepackage{anyfontsize}
\usepackage{comment}
\usepackage{soul}
\usepackage{gensymb}
\usepackage{multibib}
\usepackage[most]{tcolorbox}
\usepackage{balance}
\usepackage{footmisc}

\usepackage{listings, soul}
\usepackage[ruled,linesnumbered]{algorithm2e}
\def\BibTeX{{\rm B\kern-.05em{\sc i\kern-.025em b}\kern-.08em
    T\kern-.1667em\lower.7ex\hbox{E}\kern-.125emX}}

\usepackage{xspace}

\newboolean{showcomments}
\setboolean{showcomments}{false}         
\ifthenelse{\boolean{showcomments}}
  {\newcommand{\nb}[2]{
  \fbox{\bfseries\sffamily\scriptsize#1}
     {\sf\small$\blacktriangleright$\textit{\textcolor{red}{#2}}$\blacktriangleleft$}
   }
  }
  {\newcommand{\nb}[2]{}
   
  }

\newcommand\changed[1]{{\color{black}#1}} 
\newcommand\newnumber[1]{{\color{black}#1}} 

\begin{document}

\title{Real Faults in Deep Learning Fault Benchmarks: \\How Real Are They?}

\author{
	\href{https://orcid.org/0000-0002-1423-1083}
	{\includegraphics[scale=0.06]{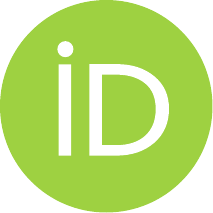}\hspace{1mm}Gunel Jahangirova} \\
	Department of Informatics \\
	King's College London \\
	London, UK \\
	\texttt{gunel.jahangirova@kcl.ac.uk} \\
	 \And
	\href{https://orcid.org/0000-0002-3037-8368}
	{\includegraphics[scale=0.06]{orcid.pdf}\hspace{1mm}Nargiz Humbatova}\\
	Software Institute\\
	Università della Svizzera italiana (USI)\\
	Lugano, Switzerland \\
	\texttt{nargiz.humbatova@usi.ch} \\
	\And
	\href{https://orcid.org/0000-0002-0140-7908}
	{\includegraphics[scale=0.06]{orcid.pdf}\hspace{1mm}Jinhan Kim} \\
	Software Institute\\
	Università della Svizzera italiana (USI)\\
	Lugano, Switzerland \\
	\texttt{jinhan.kim@usi.ch} \\
	\AND
	\href{https://orcid.org/0000-0002-0836-6993}
	{\includegraphics[scale=0.06]{orcid.pdf}\hspace{1mm}Shin Yoo} \\
	School of Computing\\
	KAIST \\
	Daejeon, Republic of Korea \\
	\texttt{shin.yoo@kaist.ac.kr} \\
		\And
	\href{https://orcid.org/0000-0003-3088-0339}
	{\includegraphics[scale=0.06]{orcid.pdf}\hspace{1mm}Paolo Tonella} \\
	Software Institute\\
	Università della Svizzera italiana (USI)\\
	Lugano, Switzerland \\
	 \texttt{paolo.tonella@usi.ch} \\
}

\maketitle

\begin{abstract}
As the adoption of Deep Learning (DL) systems continues to rise, an increasing number of approaches are being proposed to test these systems, localise faults within them, and repair those faults. The best attestation of effectiveness for such techniques is an evaluation that showcases their capability to detect, localise and fix real faults. To facilitate these evaluations, the research community has collected multiple benchmarks of real faults in DL systems. In this work, we perform a manual analysis of 490 faults from five different benchmarks and identify that 305 of them are eligible for our study. Our investigation focuses specifically on how well the bugs correspond to the sources they were extracted from, which fault types are represented, and whether the bugs are reproducible. Our findings indicate that only 15.4\% of the faults satisfy our realism conditions. Our attempts to reproduce these faults were successful only in 54\% of cases. 
\end{abstract}


\section{Introduction}



The growing popularity of deep learning (DL) systems and their widespread adoption makes ensuring that they are fault-free a task of utmost importance. In response to this challenge, in the last few years, research in the Software Engineering (SE) field has strongly focused on proposing approaches for testing, fault localisation, and repair of such systems. An obvious requirement for evaluation of these techniques is to demonstrate that they can detect, localise or fix faults in faulty DL systems. The need for faulty versions of programs has long been essential for evaluating similar approaches that target traditional software systems, the initial response to which was to apply mutation testing tools to generate versions of a program with small syntactic changes. Later, the discussion on whether the mutants are the worthy replacement for real-world faults has motivated the researchers to create large datasets of real reproducible faults such as Software-artifact Infrastructure Repository (SIR)~\cite{SIR}, Defects4J~\cite{just2014defects4j}, and BugsJS~\cite{gyimesi2019bugsjs}. 

In contrast, research on techniques for DL systems rarely relies on mutation testing tools to generate faulty DL models. Instead many publications include both a proposed approach and a dataset curated specifically to evaluate this approach~\cite{cao2022deepfd,nikanjam2021neuralint,wardat2022deepdiagnosis},
while there are also publications the sole purpose of which is to create an independent benchmark of DL faults, such as Defects4ML~\cite{morovati2023bugs} and gDefects4DL~\cite{liang2022gdefects4dl}. 

As evidenced by Defects4J~\cite{just2014defects4j} (a benchmark of 835 faults in 17 Java projects), such benchmarks can become essential in SE research, with almost all proposed approaches being evaluated on them, to the extent that there is a risk that the proposed approaches are overfitted to the benchmark. Therefore, it is imperative that the benchmarks consist of faults extracted in a systematic manner and remain unchanged with respect to their source to ensure that they are indeed representative of real-world scenarios. Moreover, the benchmarks need to be diverse in terms of both the subject programs and fault types covered.

The goal of this work is to perform a critical analysis of the benchmarks of real, reproducible faults in DL systems. Despite being relatively new, these benchmarks have already been used across multiple studies and, given the growth of the research on SE for DL, are likely to see increased adoption in the future. This study does not aim to evaluate testing techniques, fault detection methods, or localisation and repair approaches. Instead, our focus is solely on the benchmarks that have been developed to support the evaluation of such methods.
Furthermore, we concentrate specifically on faults experienced when building DL systems, not the faults in DL frameworks/libraries such as Keras, TensorFlow, PyTorch, etc. A DL fault takes place when the behaviour of the DL component is inadequate for the task at hand, meaning it is `functionally insufficient' as per ISO/PAS 21448:2019~\cite{isopas,Humbatova:2020} and the root cause of this inadequacy is due to a human error made during the development and training of the DL model. 

We first perform a systematic literature search to identify the works that contain benchmarks of reproducible real faults. We proceed with a manual analysis of each fault in the benchmarks and check whether it satisfies the following \textbf{four realism conditions}: 
(1) \textit{the benchmark reproduction code for the buggy version should match the buggy code reported in its source (such as a bug report)}; (2) \textit{the fix applied in the benchmark should match the fix reported in the source}; (3) \textit{the training data used in the benchmark must match the one mentioned in the source};
(4) \textit{the training data must be realistic}, i.e., it refers to well-known existing datasets, it was collected with a solid procedure, possibly involving manual labelling, or it was generated using a sound mathematical/statistical procedure (not just randomly).

If the bug (we use the terms bug and fault synonymously) passes all the mentioned checks, we run the bug and check whether it is indeed \textit{reproduced} and whether this reproduction is \textit{stable} across multiple runs. Lastly, based on the taxonomy of DL faults by Humbatova et al.~\cite{Humbatova:2020}, we extract the fault types that are represented in each bug and check whether these faults can be simulated with the use of a mutation tool. Our overview of 490 bugs from five different benchmarks reveals that 185 of them are unsuitable for our study (see Section \ref{sec:excl-bugs} for details). The analysis of the remaining 305 bugs shows that 
all four realism conditions are satisfied for only 47 of them. We then attempt to reproduce the faults. For the faults related to tensor problems, we do not require the realism conditions related to training data to be satisfied. This adjustment increased the number of bugs eligible for reproduction attempts to 90. Ultimately, we successfully reproduced 49 of these bugs.
The main contributions of this work are as follows:

\begin{itemize}
	\item We conduct a systematic literature review that compiles a list of works that use real DL faults in their approach/evaluation.
	
	\item We propose a set of characteristics for the analysis of the bugs to check their realism.
	
	\item
	We share a subset of the analysed faults that are realistic, reproducible and stable.
	
	\item We offer an in-depth discussion on the implications of our findings for improving bug benchmarks and mutation testing of DL systems.
	
\end{itemize}

A key focus of our analysis is \textbf{reproducibility}, which we treat as essential to the long-term value of fault benchmarks. We dedicate a specific research question to this aspect, recognising that reproducible benchmarks enable consistent evaluation, facilitate comparative studies, and ensure that research findings can be reliably built upon. In addition to their research value, we believe in the educational potential of high-quality benchmarks. Datasets that include well-documented real faults and associated fixing strategies can serve as powerful teaching tools. They can support the instruction of fault detection, debugging techniques, and model robustness evaluation, helping students and practitioners develop a deeper understanding of the challenges in building reliable deep learning systems.

We believe our work is a timely intervention in the rapidly growing field. Our goal is not to criticise ongoing efforts but to make a constructive contribution to the dialogue about best practices. By rigorously examining the benchmarks of faults, we aim to highlight areas for improvement and encourage the community to adopt higher standards that will hopefully lead to the healthy development of the research area.

\section{Fault Dataset Extraction}

\subsection{Database Search}

We performed a systematic literature search to identify the list of works that include reproducible real faults in DL systems. The search was conducted using Scopus, a large, multi-disciplinary database of peer-reviewed literature that offers an advanced search functionality. 
Our search string is designed to look for the terms related to the real bugs in any part of a paper while checking for the terms associated with DL or ML systems only in the abstract. We made this choice based on the assumption that if a paper focuses on DL or ML systems this will be brought up in the abstract. In contrast, the presence of real faults might not be the main focus of the paper but a side effect of the performed evaluation, and therefore, real faults might be not mentioned in the abstract. In addition to these search terms, we also added statements to our search string that 1) filter out the papers that do not come from a journal or conference proceedings, 2) limit the subject area to computer science, and 3) ensure that the paper is written in the English language. The final search query used with Scopus is:

\begin{tcolorbox}[colback=white,boxsep=0pt,top=2pt,bottom=2pt]
	\small
	( ( ( 
	ALL ( "real* bug*" OR "real* fault*" OR "real* defect*" OR "real* error*" ) 
	
	AND ABS ( "deep learning" OR "deep neural network" OR "machine learning" 
	) ) ) )  AND 
	
	( LIMIT-TO ( DOCTYPE , "ar" ) OR LIMIT-TO ( DOCTYPE , "cp" ) ) AND  ( LIMIT-TO ( SUBJAREA , "COMP" ) ) AND ( LIMIT-TO ( LANGUAGE , "English" ) )
\end{tcolorbox}

\subsection{Abstract and Full Text Analysis}

The application of the search string returned 282 results across diverse computer science publication venues such as "Machine Vision and
Applications", "Mobile Information Systems", "Medical Image Analysis", "Advances in Neural
Information Processing Systems", and "Neurocomputing", etc. For each obtained paper, we had to ensure that it indeed includes reproducible real faults in DL systems. This task is hard to automate and therefore we proceeded with the manual analysis of the abstract of each paper by one of the authors. Our inclusion criterion was decided to be deliberately broad in this phase, encompassing any work that proposes an approach for any software engineering activity related to DL systems. For each excluded paper, the assessor (one of the authors) wrote a summary justifying the exclusion. The majority of the papers were removed from further analysis as they focused on approaches that use DL for tasks such as fault localisation, program repair, mutation testing, testing adequacy, and oracle generation for traditional software programs. While not being directly relevant, such papers easily fulfil the requirements of our search string: they use DL-related terms and the evaluation of the proposed approaches often requires the presence of real faults. Another category of papers that we eliminated was the works focusing on real faults in DL libraries, which as mentioned before are not in the scope of our study.

As a result of the abstract analysis, we excluded 209 works out of 282, which left us with 73 papers. We downloaded and read the full text of each paper to verify whether it included a set of real faults. At this step, we again encountered papers that focused on faults in DL libraries, but understanding their scope required reading beyond the abstract. While they indeed focused on faults in DL systems (such as fault mining from StackOverflow (SO) and GitHub, as well as analysing these faults for various objectives), the majority of papers were excluded because they did not intend to be reproducible and therefore did not refer to any publicly available dataset of buggy and fixed pairs. The analysis of the full text eliminated 63 papers from consideration, resulting in a list of 10 papers left for further analysis. 

\subsection{Forward and Backward Snowballing}

To ensure that our search string did not miss any relevant studies, we manually applied both forward and backward snowballing~\cite{wohlin2014guidelines} to the identified 10 papers. Backward snowballing refers to going through the list of references of a given paper, while forward snowballing focuses on the papers that cite the given paper. We used Google Scholar for forward snowballing. Our first point of judgement was the title of the paper and we included any paper that had a DL-related term in its title. This gave us a pool of 147 titles which were reduced to 70 after removing the duplicates. We then checked how many of these titles were already covered by our search string and identified that only 48 of them were new. We performed abstract and full-text analysis for these papers as described in the previous subsection and obtained a list of 6 additional papers.

\subsection{Final Pool of Papers}

We proceeded with the analysis of our pool of final 16 papers (10 from the initial search + six from snowballing). The goal of this step was to perform the final analysis and to identify whether these papers come with a dataset suitable for the purpose of our study. For a benchmark to be deemed suitable the following conditions need to be satisfied: 

\begin{enumerate}
	
	\item The dataset should provide a source from which each fault has been extracted (such as a link to a corresponding GitHub issue commit or a unique ID for an SO post that raises/discusses this fault). This condition must be strictly fulfilled because, without the source of a fault, it is impossible to check whether the version of the fault specified in a fault benchmark matches the original version. 
	
	\item For each fault in the benchmark, the source code and training dataset for the faulty and the fixed version need to be provided.
	The presence of the fixed version is required because, in the absence of such a fix, it is not possible to claim that the reported issue is indeed a fault and that it is possible to fix it. 
	Note that the fixed version also needs to come from the source of the fault to be considered realistic. If it is from GitHub, then the fix should be present in the issue discussion or in the committed code. If the fault originates from SO, then the fix should be provided in one of the answers to the post. 
	The cases when fixed versions come from the fixes proposed in the papers that, for example, target program repair are not acceptable, as we do not know whether this is really a ground truth fix for a given fault (e.g., it might have been crafted by the paper authors, who are biased by the technique they developed).
\end{enumerate}

In our pool of 16 papers, one of the papers~\cite{tensfa_journal} is a journal extension of a previously published conference paper~\cite{tensfa_conf} and both of them are using the same dataset of real faults. Similarly, the work by Kim et al.~\cite{kim2023repairing} uses a subset of faults from the dataset introduced by Cao et al.~\cite{cao2022deepfd}. For such cases, we keep only the paper that introduces the original dataset. 

\subsection{Excluded Datasets}
The work by Nikanjam et al.~\cite{nikanjam2021neuralint} introduces \textit{NeuraLint}, a model-based fault detection approach for DL programs that uses meta-modeling and graph transformations. The authors evaluate \textit{NeuraLint} on 34 real-world DL models extracted from SO posts and GitHub repositories. However, our inspection of its replication package identified that the source code is provided only for the buggy version and not for the fixed version. 
Zhang et al.~\cite{zhang2020debar} propose a static analysis approach named \textit{DEBAR} for detecting numerical bugs in DNNs based on abstract interpretation. \textit{DEBAR} is evaluated on DNNs
with confirmed bugs 
(collected from works by Zhang et al.~\cite{zhang2018issta} and Odena et al.~\cite{odena2019tensorfuzz})  and real-world
neural architectures collected from TensorFlow Models GitHub repository~\cite{tfmodels}. Similarly to \textit{NeuraLint}~\cite{nikanjam2021neuralint}, the replication package of 
\textit{DEBAR} contains files only for the buggy DNNs, and therefore it also had to be excluded from our study.

Yan et al.~\cite{yan2021exposing} proposes \textit{GRIST}, a technique for generation of inputs that expose numerical bugs in DL programs. The evaluation of \textit{GRIST} was performed on 63 real-world DL programs that authors collected from GitHub. However, this evaluation does not rely on any ground truth fixed DL programs. Instead, the authors report the bugs detected by \textit{GRIST} in the issue repositories of DL programs and expect the developers to fix them. This leads to the exclusion of this work, which then creates a chain reaction of also excluding the work by Li et al.~\cite{li2023reliability}, as this work indicates that it uses the \textit{GRIST} dataset for its evaluation.

The work by Liu et al.~\cite{liu2021detecting} has a specific goal of detecting TensorFlow bugs in industrial systems and proposes a constraint-based approach named \textit{ShapeTracer}. As the subject programs come from industry, the evaluation dataset, consisting of 60 TensorFlow programs, could not be made available in their replication package. 

The work by Zhang et al.~\cite{zhang2021autotrainer} proposes \textit{AutoTrainer}, a rule-based automatic repair tool that supports detecting and fixing five commonly seen training problems. The evaluation of \textit{AutoTrainer} is performed on 495 models which as reported by the authors were \textit{``collected from reported buggy models on GitHub, SO, existing papers and personal blogs, and some of them are gathered from machine learning experts within our organisation''}. However, no further details are provided on this, preventing us from accessing the source for each fault. 

The work by Wardat et al.~\cite{wardat2022deepdiagnosis} introduces \textit{DeepDiagnosis}, a debugging approach that localises faults, reports error symptoms and suggests fixes for DNN programs. To evaluate \textit{DeepDiagnosis}, the authors employed a dataset of 53 real-world DNNs from GitHub and SO, and 391 from \textit{AutoTrainer}. While the ones from \textit{AutoTrainer} are excluded for the reasons mentioned before, we proceeded to look into the 53 DNNs from GitHub and SO. Unfortunately, the replication package of \textit{DeepDiagnosis} only provides the IDs of posts from SO, while providing no point of reference for the GitHub issues. Moreover, there is no source code provided for the faulty and ground truth versions of each analysed bug. 

The work by Li et al.~\cite{li2021cleanml} performs an empirical study to investigate the impact of existing data cleaning on ML classification tasks. For this purpose, the authors use 14 real-world datasets with real errors. The replication package for this work contains a pdf document that provides a high-level textual description of the errors. However, there are no pairs of faulty and cleaned ground truth datasets provided, making it impossible for us to explore these datasets as example of real faults in training data. 
The work by Liang et al.~\cite{liang2022gdefects4dl} presents gDefects4DL, a dataset of 64 bugs. Along with the buggy/fixed versions, the authors claim to provide an isolated environment to replicate each bug. The paper contains links to both a cloud service using which the bugs can be downloaded and a web page that can be used to navigate the dataset. However, both links are no longer active. We have contacted the authors of the benchmark and they have confirmed the problem with the links. Our request to provide access to the dataset did not receive a response at the time of submission.

Overall, after removing the works that did not satisfy the criteria for the benchmark of real reproducible faults or were not available, we were left with five benchmarks to perform our realism analysis on.

\section{List of Fault Benchmarks}

The five benchmarks that survived our multi-step elimination process are: \textit{DeepLocalize}~\cite{wardat2021deeplocalize}, \textit{DeepFD}~\cite{cao2022deepfd}, \textit{Defects4ML}~\cite{morovati2023bugs}, \textit{SFData}~\cite{tensfa_conf,tensfa_journal}, and the dataset provided in the study by \textit{Zhang et al.}~\cite{zhang2018issta}. In Table~\ref{tab:benchmark_list}, we provide some brief information on these benchmarks such as their publication venue(s), main purpose, overall number of bugs and in brackets the number of analysed bugs 
(the reasons why we excluded some bugs from the analysis are detailed later, in Section~\ref{sec:excl-bugs}). As the table shows, all the benchmarks have been published at prestigious SE venues and contain a significant number of bugs. The fact that all five datasets originate from SE venues reflects the current state of research, where the creation of fault benchmarks is largely driven by the SE community. These benchmarks are typically intended to support the evaluation of fault detection, localisation, and repair tools—topics that are central to software engineering research.

\begin{table}[ht]
	\caption{List of analysed benchmarks}\label{tab:benchmark_list}
	\centering
	\begin{tabular}{c|c|c|c}
		\toprule
		\textbf{Dataset} & \textbf{Venue} & \textbf{Goal} & \textbf{\#Bugs} 
		\\
		\midrule
		TFBugs2018~\cite{zhang2018issta} & ISSTA'18 (C) & Fault Analysis & 151 (21)\\
		DeepLocalize~\cite{wardat2021deeplocalize} & ICSE'21 (C) & Fault Localisation &  41 (38)\\
		DeepFD~\cite{cao2022deepfd} & ICSE'22 (C) & Fault Localisation & 52 (36)\\
		Defects4ML~\cite{morovati2023bugs} & EMSE'23 (J)& Fault Dataset & 100 (68) \\
		& ISSRE'21 (C) &   & \\      
		\multirow{-2}{*}{ SFData~\cite{tensfa_conf,tensfa_journal}}	 & 	IST'22 (J) &  	\multirow{-2}{*}{ Fault Repair} 		 &  	\multirow{-2}{*}{ 146 (142)} 	\\
		\bottomrule
	\end{tabular}
\end{table}

The work by \textbf{\textit{Zhang et al.}}~\cite{zhang2018issta} focuses on studying the characteristics of faults in DL programs built on top of the TensorFlow framework. The authors investigate the symptoms and root causes of the faults, as well as the main challenges associated with detecting and localising them. This analysis has resulted in a benchmark of 151 faults (75 from GitHub and 76 from SO) classified across seven root causes. We refer to this dataset as TFBugs2018.

\textbf{\textit{DeepLocalize}}~\cite{wardat2021deeplocalize} is a fault localisation tool that automatically determines whether a model under test is faulty and identifies the root causes of the bug detected. It does so by adding a callback mechanism to the model's training process and analysing the historic trends in values propagated between layers. The evaluation of \textit{DeepLocalize} is performed on 41 bugs, 11 coming from GitHub and 30 from SO.

\textbf{\textit{DeepFD}}~\cite{cao2022deepfd} is a fault diagnosis and localisation framework which maps the fault localisation task to a learning problem instead of employing a rule-based approach as in \textit{DeepLocalize}. The evaluation of \textit{DeepFD} is performed on 52 real faults (5 from GitHub and 47 from SO). As the starting point for mining the faults, the authors indicate previous studies that focused on analysing types of DL faults without reproducing them~\cite{Humbatova:2020,islam2020repairing} as well as the benchmark of \textit{DeepLocalize}. A closer look into the benchmark reveals that 34 out of its 52 faults come from \textit{DeepLocalize}. However, we checked whether the files provided for faulty and fixed versions are identical and this was not the case. Our understanding is that while the authors used the same SO posts and GitHub commits as sources of the faults, they performed the process of reproducing the bug independently.

\textbf{\textit{SFData}}~\cite{tensfa_conf,tensfa_journal} is a fault benchmark that has been collected as part of the works by Wu et al., who proposed \textit{Tensfa}~\cite{tensfa_conf} and  \textit{Tensfa2}~\cite{tensfa_journal}. Both approaches aim to automatically repair tensor shape faults in DL programs. The 146 faults in \textit{SFData} all come from SO and are collected by the authors to evaluate the proposed approaches. 

The work by Morovati et al.~\cite{morovati2023bugs} focuses on the challenges attributed to the process of extracting reproducible and verifiable faults in DL-based systems. The main contribution of this study is a \textbf{\textit{Defects4ML}} benchmark of 100 pairs of the faulty and fixed programs originating from GitHub or SO that were either mined by the authors or extracted from existing literature~\cite{islam2020repairing,Humbatova:2020,wardat2021deeplocalize,zhang2018issta,nikanjam2021neuralint}. In their selection process, the authors considered fault types that fall under the main categories of the taxonomy of real faults in DL systems~\cite{Humbatova:2020}. Out of the five benchmarks analysed in our study, \textit{Defects4ML} is the only one that has been published in a paper that specifically aims to create a DL fault benchmark, while all the others were generated to be used in the evaluation of the approach proposed in their original publications.

\section{Criteria for Fault Realism}
\label{sec:criteria}
The primary goal of this work is to systematically assess whether the faults included in deep learning (DL) real-fault benchmarks can genuinely be considered real. In the broader software engineering literature, numerous benchmarks have been developed to collect real faults across different languages and domains~\cite{SIR,just2014defects4j,bugsjs,bugswarm,manybugs}. These benchmarks are typically grounded in the intuitive understanding that real faults are those that occurred during  development processes. However, they rarely provide a formal definition of what precisely constitutes a real fault.

Despite this lack of formalisation, several benchmarks convey a shared conceptual understanding of real faults. For instance, the \textbf{SIR} benchmark~\cite{SIR} refers to “\textit{naturally occurring faults}”, emphasising that such faults “\textit{are costly to locate and typically cannot be found in large numbers, but they represent actual events}”. This highlights that realism is closely tied to faults being genuine historical incidents. Similarly, \textbf{Defects4J}~\cite{just2014defects4j} defines real bugs as “\textit{bugs fixed by a developer}”, directly associating realism with the presence of a fix authored by a developer. In the \textbf{BugsJS} benchmark~\cite{bugsjs}, realism is ensured by the process used to collect faults: the authors “\textit{identify existing bugs from the programmes’ version control histories, and collect the real fixes provided by developers}”. Likewise, the \textbf{BugSwarm} dataset~\cite{bugswarm} explicitly states that “\textit{natural faults, from real programmers, would provide more realism}”, thereby tying the notion of realism to the origin of a fault.
The \textbf{ManyBugs} benchmark~\cite{manybugs} takes a systematic approach to realism by ensuring that each defect “\textit{has at least one corresponding patch and test case written by an original developer}”, and that the defects “\textit{were captured systematically through version control repositories}”. 

Across all these works, real faults are consistently collected from well-documented repositories by identifying fault-fixing commits, extracting the version preceding the fix to obtain the buggy code, and applying the corresponding patch to reconstruct the fixed version. This shared understanding not only shapes how realism is defined, but also guides how faults are collected—typically from version control histories that faithfully capture authentic development activity. This process ensures that both the buggy and fixed code in a benchmark accurately reflect the versions used during real development.

Based on this observation, we introduce our first two realism criteria for evaluating deep learning fault benchmarks:

\noindent \textbf{Realism Criterion 1:} \textit{\changed{The benchmark reproduction code for the buggy version should match the buggy code reported in its source.}} 

\noindent \textbf{Realism Criterion 2:} \textit{The fix applied in the benchmark should match the fix reported in the source.} 

\begin{figure}[htbp]
    \centering
    \includegraphics[width=0.75\linewidth]{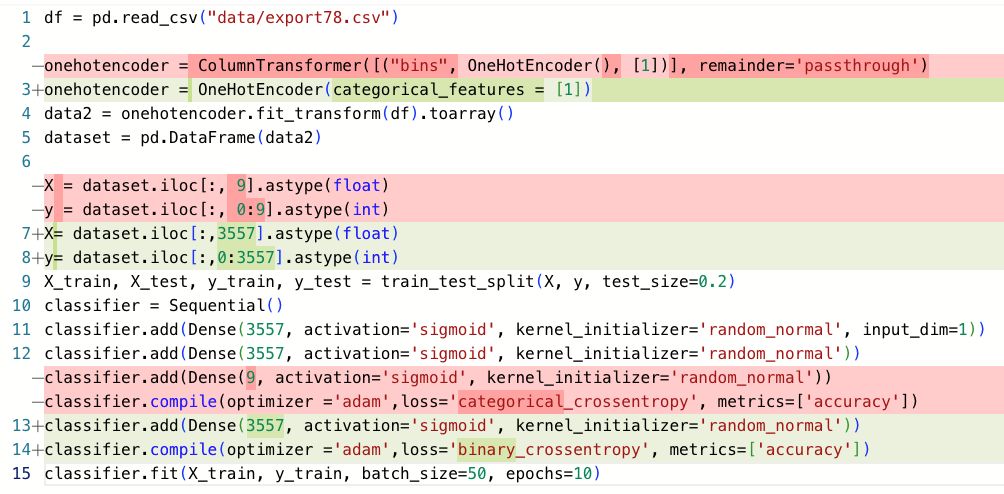}
    \caption{An example of a case where the buggy version in the benchmark does not correspond to the buggy version provided in StackOverflow post 58609115.}
    \label{fig:buggy_corresponds}
\end{figure}

Figure~\ref{fig:buggy_corresponds} illustrates a violation of Realism Criterion~1. The figure presents the diff between the code provided in StackOverflow post 58609115, which serves as the basis for a bug in the DeepFD benchmark, and the buggy version included in the benchmark. \changed{It can be seen that the authors of the benchmark replaced the original ColumnTransformer-based encoding with a direct OneHotEncoder configuration, changed the selected input feature while preserving the input dimensionality (one feature), significantly increased the output dimensionality (from 9 to 3557 features), replaced the original 9-unit output layer with a 3557-unit output layer, and changed the loss function from '\textit{categorical\_crossentropy}' to '\textit{binary\_crossentropy}'.}
These modifications affect core aspects of the system and therefore the benchmarked version no longer corresponds to the original faulty implementation, breaking traceability to the real-world issue.

A violation of Realism Criterion 2 can be observed in the DeepFD benchmark fault derived from StackOverflow post 59325381. 
In the original post, the author reports low accuracy (10\%) when training a CNN model for handwritten digit classification in Keras. The accepted answer identifies a likely typo (rescaling the training labels instead of the test data) and states that correcting this issue increases accuracy to 95\%, which the original poster confirms by accepting the answer. However, in the benchmark, the fix includes not only the correction suggested in the accepted answer but also several additional modifications, including changing activation functions of the last two layers (from '\textit{ReLU}' to '\textit{SELU}' and from '\textit{Softmax}' to '\textit{Sigmoid}') and altering the batch size. These changes are not mentioned in the accepted solution and significantly affect the model behaviour. In particular, replacing a '\textit{Softmax}' output layer with a '\textit{Sigmoid}' activation is inappropriate for a multi-class MNIST classification task, as this activation function is typically used for binary classification settings. Since the applied fix does not correspond to the solution agreed upon in the original StackOverflow exchange, the benchmarked fix cannot be considered a faithful reconstruction of the real-world resolution. 
 
In contrast to traditional software systems, deep learning (DL) models derive much of their behaviour not only from their architecture and code but also from the data on which they are trained. This makes training data a central component of the system's functionality, and thus a critical factor when evaluating the realism of faults. Unlike conventional programs, where faults are typically tied to logic or control flow in the code, faults in DL systems might emerge from unexpected interactions between the model code and the data, or from data-dependent training dynamics. As a result, altering the training data used in the original development scenario may substantially affect the manifestation of the fault. Therefore, to ensure the fault remains faithful to its original context, it is essential that benchmarks use the same dataset as the one reported in the source. This motivates our next criterion:

\noindent \textbf{Realism Criterion 3:} \textit{The training data used in the benchmark must match the one mentioned in the source.} 

An example of a case when this realism criterion is violated can be observed in DeepLocalize benchmark fault that has been extracted from StackOverflow post 48251943. In the post the author describes a regression task trained on a dataset loaded from "Phenols-toxicity.csv". The reported issue is that the model predicts zero whenever the true target value is negative. The accepted answer correctly explains that this behaviour is caused by using a ReLU activation function in the output layer, which by definition truncates negative outputs to zero. The fault therefore lies in an inappropriate architectural choice for a regression problem that requires unconstrained continuous outputs.

In the constructed benchmark, however, the original dataset is not used, nor is there any attempt to retrieve or approximate "Phenols-toxicity.csv". Instead, the authors extract the list of expected target values shown in the post (i.e. values that were provided merely to illustrate the prediction error) and treat them as the entire dataset. They place these values into a CSV file, divide them sequentially into eight columns, and use this artificially reshaped file as training data.

This fundamentally misrepresents the original scenario. The StackOverflow post concerns a supervised regression task with input features and corresponding targets. By replacing the original dataset with a reshaped list of target values, the benchmark removes the feature space entirely and introduces artificial structure that has no basis in the original problem. As a result, the nature of the task is altered, and the reproduced setting no longer faithfully reflects the real debugging scenario described in the post.

In addition to the training data being required to correspond to the source from which the fault has been extracted, it is important that the dataset itself reflects realistic data, i.e. data that could plausibly be encountered in a real-world application scenario. Artificial or overly simplified datasets may fail to capture the complexity and variability that typically lead to faults in deployed systems. To ensure the credibility of the benchmark, we introduce our last criteria:

\noindent \textbf{Realism Criterion 4:} \textit{The training data must be realistic.} 

\begin{figure}[htbp]
    \centering
    \includegraphics[width=0.7\linewidth]{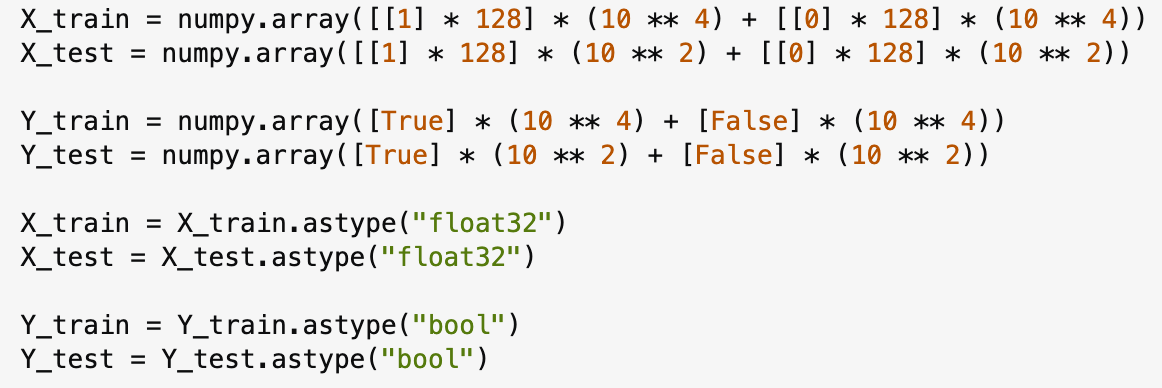}
    \caption{An example of non-representative artificial training data from StackOverflow post 31880720 used in DeepFD benchmark.}
    \label{fig:synthetic_data_rc4}
\end{figure}

Simplistic synthetic data that does not correspond to any real-world training scenario can be observed in the DeepFD dataset fault extracted from StackOverflow post 31880720. In the original post, the author states that they are attempting to build a model for the MNIST dataset. However, the provided code (see Figure \ref{fig:synthetic_data_rc4}) manually constructs artificial training data rather than using the actual MNIST dataset. The DeepFD benchmark reproduces the fault with this exact same artificial data.

In this case, although the benchmark reproduces the data construction shown in the original post, the training data is artificially generated and extremely limited in size and complexity. Such simplified, synthetic inputs do not resemble realistic data distributions encountered in practical machine learning scenarios. As a result, models trained on this data are not representative of real-world systems, and faults derived from them lack practical relevance, since they do not reflect failures that would meaningfully occur in authentic training environments.

\section{Analysis of Fault Benchmarks}

\subsection{Research Questions}

The \textbf{goal} of our study is to critically evaluate the reproducibility and realism of fault benchmarks used in the research area of ``Software Engineering for DL'', as well as provide some insights into how representative they are of DL faults. To this extent, we have performed a set of analyses and experiments to answer the following research questions:  

\textbf{RQ1 [Correspondence to Source]}: \textit{Do the faulty versions in the benchmarks and their fixes correspond to the source from which the fault was extracted?}

This RQ is important to understand the realism of the faults in the considered benchmarks. A DL fault can be reliably considered a real DL fault if both the faulty and the fixed versions of the code match an existing source, where developers reported them.

\textbf{RQ2 [Training Dataset]}: \textit{Does the training data used in the benchmarks correspond to the source from which it was extracted and is it realistic?} 

This RQ is also about the realism of the faults in the considered benchmarks, but it focuses on the training data instead of the source code. In fact, a real DL fault can be reproduced faithfully only if the training process is carried out on the same training dataset used originally by developers.

\textbf{RQ3 [Fault Types]}: \textit{What types of faults do the benchmarks contain and what is the frequency of each fault type?}

This RQ is about the representativeness of the DL faults that appear in the considered benchmarks. We want to see whether faults are distributed evenly across fault types or present a skewed distribution, which would make them less representative.

\textbf{RQ4 [Similarity to Artificial Faults]}: \textit{Can the faulty models in the benchmarks be generated by artificially injecting faults using existing mutation operators?}

This RQ is about the possibility of simulating real DL faults using DL mutation tools. If confirmed, such a possibility would enable evaluation campaigns on larger benchmarks, obtained from mutation tools.

\textbf{RQ5 [Fault Reproducibility]}: \textit{Are the faults in the benchmarks reproducible and stable across multiple runs?}

This is a key RQ, as stable reproducibility of the faults in the benchmarks is a fundamental prerequisite for their practical usage. We want to assess both the technical aspects associated with fault reproduction (including the required libraries, operating system, and resources), but also its stability (e.g., whether flaky, unstable results are obtained in multiple runs).

\subsection{Experimental Procedure}

\subsubsection{RQ1 [Correspondence to Source]} To extract the source code of the faulty and fixed code pairs in the benchmarks, we first locate the links to the replication packages in the corresponding papers. The replication packages usually contain spreadsheets listing all the faults with their source indicated and the pairs of source code files for each fault. 
For a fault to be considered ``real'', its faulty version in the \changed{benchmark reproduction code} must correspond to the source it was extracted from. 

For each item in the list, we follow the links to SO posts or GitHub commits. In GitHub, the link takes us to a commit that includes changes in the files associated with the fault. For each file, we extract the source code before the commit. In SO, we copy the code provided in the post. 
We then generate a diff between the code obtained from GitHub or SO and the faulty version from the benchmark, and we manually analyse it. If the changes in the diff are only pieces of code added for the purpose of debugging (e.g., \textit{print} statements; different verbosity levels in \textit{model.fit} statement) or different syntax that leads to the same semantics (e.g., indicating the activation as the parameter of the layer instead of adding it as a separate layer or declaring the hyperparameter value as a separate variable instead of passing its value directly to the \textit{model.compile} statement), we ignore them. However, if the diff contains alterations to the code that lead to a different model structure, hyperparameter value or training process, we mark the fault as not corresponding to its source.  It should be noted that sometimes in SO
not the whole code is available, but only the model structure. In such cases, we require only the available code not to be changed, as this part is reported to be the cause of the bug. The remaining missing parts of the code can be adjusted by the authors of the benchmark as they see fit, since there is no source that we can use to check the correspondence to.

Our next step is to check whether the fix applied to the faulty model also corresponds to the source of the fault. In GitHub, we extract the code after the commit and compare it to the fixed source code provided in the benchmark. In SO, very often there is no fixed source code available and the fixes are provided as textual suggestions. Whether the provided suggestion indeed solves the reported problem is usually indicated by the post author marking the answer with the suggestion as ``accepted''. Therefore, the first point of our analysis is to check whether an SO post has an accepted answer. 
If this is not the case, then it is not possible to claim that there exists a ground truth fix for the given fault. 
However, we still check if any of the ``non-accepted'' answers correspond to the fix applied in the benchmark, as they might still be fixing the reported problem. 
Note that as there is not necessarily the source code of the fix in SO, we read the answers and manually check whether the suggestions provided in them are indeed applied in the fixed files in the benchmark. Given that the answers are written in free text format, their direct application to the source code might be prone to subjectivity. To ensure we keep this process systematic, we apply the following rules: 

\begin{enumerate}
	\item If the answer contains a list of suggestions and the fix consists of a subset of these suggestions (but not all of them), we mark the fix as corresponding to the source. However, if a part of the fix does not correspond to any suggestion in the answers to the post, it is classified as not corresponding to the source.
	\item If the suggestion in the answer is vague, for example, it suggests changing a hyperparameter value, but does not provide the exact value to change to, the fixes that apply any change in the value are deemed as corresponding to the source.
\end{enumerate}

\subsubsection {RQ2 [Training Dataset]} For each fault in the benchmark we manually check whether the training dataset used in the benchmark is the same as the one indicated in its source (GitHub or SO). For both GitHub and SO, the employed dataset can be identified in the parts of code that load the training data. In SO, sometimes the datasets are just mentioned in the text of the post and not in the code. 
Once we check that the datasets are the same, we then proceed with analysing whether a real or a fake dataset has been used. We consider a dataset \textit{real} if it is a known existing dataset used previously for a learning task or if it is a meaningful data generated using some automation (such as data generated using mathematical formulas for a regression task). In contrast, a dataset is considered \textit{fake} if it does not refer to any existing dataset or meaningful data (e.g., it just contains randomly generated values). 
If a dataset is real, we record its name. If the dataset is fake, we further analyse the size of the dataset and whether it is generated randomly or not. 

\subsubsection {RQ3  [Fault Types]} 
In this RQ, we aim to identify and analyse types of the faults that affect real faulty models through the applied fixes. As the reference for the list of possible fault types in DL systems, we use the taxonomy by Humbatova et al.~\cite{Humbatova:2020}. This taxonomy was created by manually analysing 1,059 SO posts and GitHub issues and interviewing 20 professional developers. The taxonomy was further validated by surveying with another set of 21 developers. It is organised into five top-level categories (three of which are divided into inner subcategories) and contains 92 types of faults. 

We manually analyse the diffs between the buggy and fixed source code as provided in the benchmark and list the fault types that comprise each diff. The diffs that contain changes to multiple elements of the DL model (e.g. changing both the activation function for the last layer and the loss function for the model) are assigned with multiple fault types. If the same fault type takes place in different parts of the DL model (e.g., activation is changed for multiple layers), we list this fault type multiple times. 

\subsubsection {RQ4 [Similarity to Artificial Faults]} 

To extract the list of fault types that can be simulated by injecting artificial faults we use the DeepCrime~\cite{humbatova2023deepcrime,HumbatovaJT21} mutation testing tool. We choose DeepCrime as it is the only available tool that introduces changes to the source code of the DL model before training (instead of altering weights and biases of an already trained model \cite{hu2019deepmutationpp}). DeepCrime has 24 mutation operators extracted from 793 real faults in DL systems and has been implemented for the Keras platform. These mutation operators introduce changes to the training data, hyperparameters, activation function, regularisation, weights initialisation, loss function, optimisation function and validation process. 

However, DeepCrime injects only single-order mutants, i.e., it can change either the activation function or the loss function, but not both of them at the same time. In contrast, the faults in the benchmarks can contain changes to multiple elements. We use the analysis performed in RQ4 that extracts the list of fault types present in each fault of the benchmarks. We then manually compare each of these fault types (that correspond to single-order mutants) against the 24 mutation operators of DeepCrime and check whether each of them could have been injected by DeepCrime. In fact, applying DeepCrime multiple times could lead to generating the composite faults present in the benchmark.
As a result of this analysis, for a composite fault $X$ that contains $n$ changed elements $(X_1, X_2, ..., X_n)$, we will obtain a list of size $n$ where each $i$-th element has a boolean value indicating whether the change to element $X_i$ can be simulated by DeepCrime. We will refer to the elements $X_i$ as \textit{fault components}.

\subsubsection {RQ5 [Fault Reproducibility]} 

For each analysed fault, we identify whether it satisfies our four conditions for realism. If that is the case, then we attempt to reproduce the bug. It should be noted that as SFData has a specific focus on tensor shape faults and the usage of realistic training data is not relevant to the reproduction of such faults. Therefore, we remove the two realism conditions on training data for the bugs from this dataset.

The reproducibility of the bugs is highly dependent on the versions of Python and of the frameworks/libraries. For TFBugs2018 and SFData no versions are provided. For DeepLocalize and DeepFD the versions to run the proposed tools are indicated in their GitHub repositories. Our attempts to install these versions were unsuccessful, as these are old versions not supported by many of the modern machines. Therefore, for these four datasets, we decided to use the recent stable versions of TensorFlow and Python. In fact, in order for the faults in these benchmarks to be useful for the evaluation of newly proposed approaches, they should be reproducible with the recent versions. For Defects4ML the exact versions are specified for each bug in the dataset and we use those versions. 

To reproduce the faults, we run all buggy and fixed versions 20 times. We allowed each run to complete, even if the fix resulted in significantly longer execution times (for example, by increasing the number of training epochs). Our criteria to consider the bugs reproduced varies according to the reported symptom. If a symptom is a crash/failure, then we expect the buggy version to produce a failure message, while the fixed version should not. We verify that the crash consistently occurs in 20 repeated runs of the buggy program. Likewise, for the fixed versions, we confirm that the program complete successfully in all 20 executions. 

In cases where the fault affected performance rather than causing a crash, we ensured that the minimum performance observed in the fixed version was not lower than the maximum performance observed in the buggy version. Additionally, we performed a statistical test comparing the two sets of 20 performance measurements (i.e. accuracy or loss) to assess whether a fix led to a statistically significant improvement. We used generalised linear model (GLM) \cite{Nelder:1972}) and checked if the effect size for an improvement in performance metric is ``non-negligible'' (using Cohen's $d$ \cite{Kelley:2012}). A fault was deemed reproducible only if all of these checks were satisfied.

\subsection {Manual Analysis}

As indicated in the experimental procedure, our first four RQs require careful manual analysis by a human. To make the approach systematic, we identified the information to be extracted for each RQ and organised it into spreadsheet columns, which are available as part of our replication package \cite{replpackage}.

\begin{figure}[htbp]
	\centering
	\includegraphics[width=\linewidth]{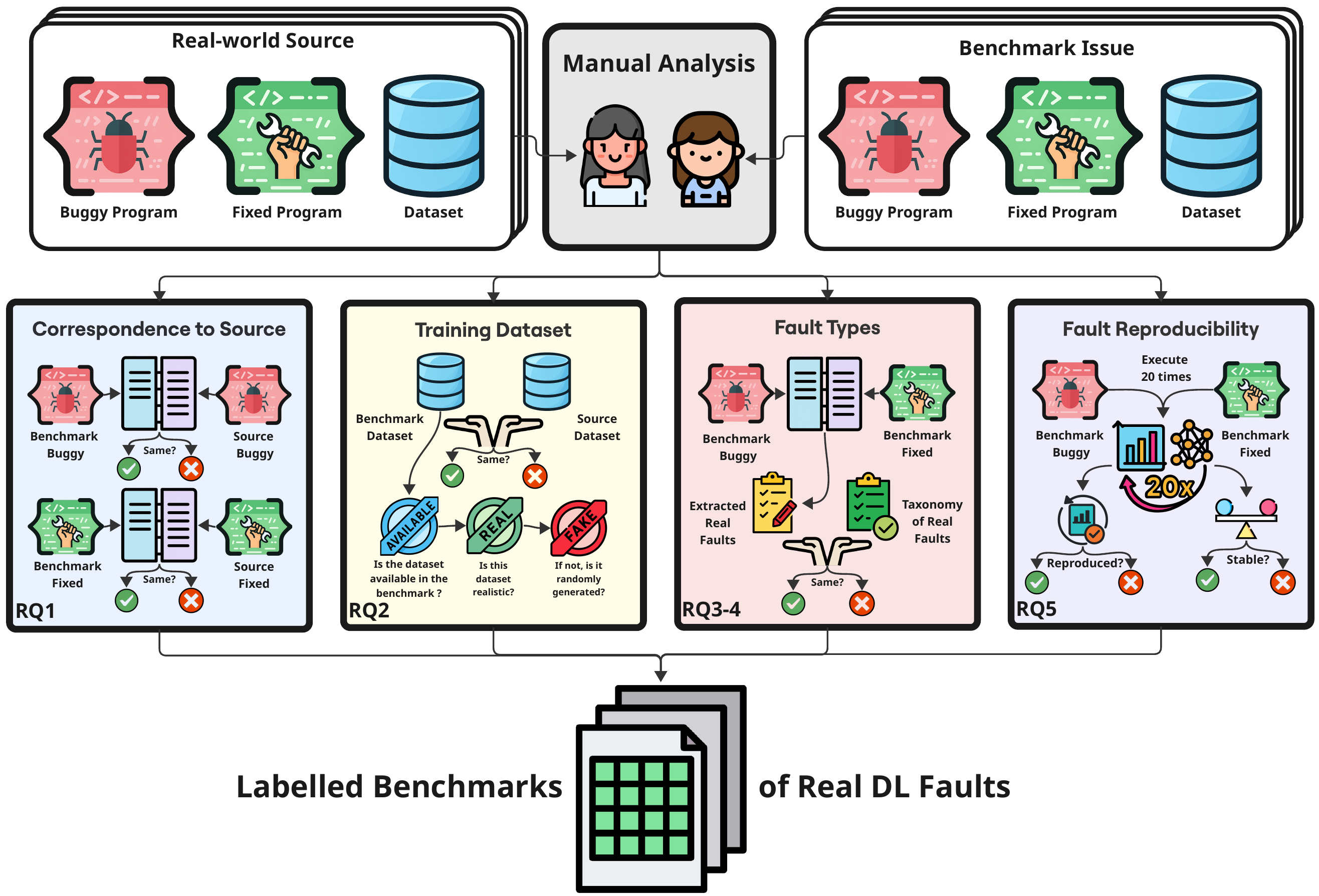}
	\caption{Manual analysis of the benchmarks.}
	\label{fig:manual_an}
\end{figure}

Figure~\ref{fig:manual_an} illustrates the core idea behind our analysis, including the main evaluation components and criteria. The primary focus of the analysis was to assess whether the representation of issues in the benchmarks corresponds to their real-world sources. In particular, we examined whether the buggy version, as well as the applied fixes, were accurately recreated from the corresponding SO discussions or Git commits (RQ1).

Since the dataset and its preprocessing may significantly influence the performance of a model trained on it, we investigated whether the datasets used in the benchmarks are available and whether they match those used in the original source issues. We further analysed whether the datasets used in the benchmarks are realistic or generated using random number generators (RQ2).

In the next stage, we analysed and documented all fixes applied to each issue and compared the corresponding code changes with the existing classification of real DL faults~\cite{Humbatova:2020} (RQ3–RQ4). Additionally, we compared the fixed faults with existing mutation operators to determine whether these faults can be reproduced by state-of-the-art mutation tools for DL systems~\cite{hu2019deepmutationpp,HumbatovaJT21}.

Finally, to conclude the analysis, we replicated the issues that passed the realism checks (Section~\ref{sec:criteria}), i.e. those that can be considered replications of real faults, to verify whether the changes introduced to the buggy version resolve the reported issues, such as execution crashes or suboptimal performance. We repeated each replication 20 times per issue to assess whether failures and fixes manifest consistently and are not merely the result of randomness in the training process.

Two of the authors independently analysed all benchmarks. The results were then compared, and disagreements were resolved during periodic meetings. Most disagreements arose from grey areas concerning (1) how to translate free-text fix suggestions from SO into concrete code changes, (2) when to classify a dataset as real or fake, and (3) how to handle composite bugs. Discussing these cases led to more precise rules and definitions, which are already described in the experimental procedure for the corresponding RQs.

\subsection{Excluded Bugs} \label{sec:excl-bugs}

A closer investigation of the benchmarks revealed that not all bugs could be analysed as part of our study. The value in brackets in column `\textit{\#Bugs}' of Table~\ref{tab:benchmark_list} reports the number of bugs that were kept for our analysis.
In DeepLocalize, DeepFD and SFData, overall \newnumber{22} bugs were excluded. The most common reasons for the exclusion of bugs were as follows: 1) the bugs referred to a GitHub link that was no longer valid, 2) the bugs were extracted from SO posts with no code or with no responses, 3) the bugs were extracted in a way that misrepresented the SO post, \newnumber{and 4) buggy and/or repaired versions of the program missing from the benchmark}.

We have excluded \newnumber{32} bugs out of 100 in Defects4ML. In seven of these cases there was no actual bug represented, as the buggy versions were either fully identical to the fixed one or the changes included only variable renaming or added print statements. In one case the files in the benchmark did not correspond to the code in the source at all, while in two cases the source of the bugs was not indicated. Moreover, the benchmark contained three duplicate entries of other issues from the benchmark. The remaining of the excluded bugs were either referring to simple `how to' questions or to  bugs not in a DL system (such as general programming errors or changes not related to DL).

Lastly, for TFBugs2018 we had to exclude all the bugs originating from GitHub and 3 bugs from SO as there were no links to corresponding sources available. A careful analysis of the remaining 73 bugs revealed further reasons for exclusion: `how to' questions (20), issues related to APIs and framework versions/compatibility (18) and bugs not in DL systems (\newnumber{8}). In 3 bugs, the benchmark entry did not bear similarities to the source. Finally, we found 3 duplicates of  existing issues from the same dataset.

\section{Results}
\subsection{RQ1 [Correspondence to Source]}

Table~\ref{tab:rq1_results} reports results on the correspondence of the faults in the benchmark and their fixes to the sources they were extracted from. Column `\textit{\#Analysed bugs}' reports the number of analysed faults, and column `\textit{Buggy m.}' reports the number of cases when the buggy code in the dataset fully corresponds to its source. Similarly, column `\textit{Fix m.}' reports the number of instances in which the fixed version in the benchmark matches the fix provided in the source. The column `\textit{Buggy \& fix m.}' combines the two conditions and reports the number of cases when both the buggy version and the fixed were extracted from their sources without introducing any additional changes.

\begin{table*}[ht]
	\caption{Results on the correspondence of the faults in each benchmark}
	\label{tab:rq1_results}
	
	\centering
	\resizebox{\linewidth}{!}{%
		\begin{tabular}{c|c|c|c|c|c|c|c}
			\toprule
			\textbf{Dataset} & 
			\textbf{\#Bugs} &
			\textbf{Buggy m.} &
			\textbf{Fix m.} & 
			\textbf{\makecell{Buggy \& \\ fix m.}} &
			\textbf{TD m.} & 
			\textbf{TD real} & 
			\textbf{\makecell{Buggy \& \\ fix m. \& \\ Real TD m.}}
			\\
			\midrule
			TFBugs2018 & 21 & 16 (76.2\%) & 19 (90.5\%) & 16 (76.2\%) & 11 (52.4\%) & 9 (42.9\%) & 4 (19.0\%)\\ 
			DeepLocalize & 38 & 23 (60.5\%) & 21 (55.3\%) & 13 (34.2\%) & 24 (63.2\%)& 29 (76.3\%)& 11 (28.9\%)\\
			DeepFD & 36 & 14 (38.9\%) & 14 (38.9\%) & 6 (16.7\%) & 29 (80.6\%)& 22 (61.1\%) & 2 (5.6\%)\\
			Defects4ML & 68 & 58 (85.3\%) & 54 (79.4	\%) & 48 (70.6\%) & 53 (77.9\%) & 47  (69.1\%) & 29 (42.60\%)\\
			SFData & 142 & 50 (35.2\%) & 114 (80.3\%) & 44 (31.0\%) & 102 (71.8\%) & 23 (16.2\%) & 1 (0.7\%)\\  
			\midrule
			Overall & 305 & 161 (52.8\%) & 222 (72.8\%) & 127 (41.6\%) & 219 (71.8\%) & 130 (42.6\%) & 47 (15.4\%)\\ 
			\bottomrule
		\end{tabular}
	}
\end{table*}

For \textbf{DeepLocalize}, in \newnumber{23} out of \newnumber{38} cases, the buggy version in the benchmark matched the code reported in its source. A closer inspection of the mismatches reveals that in \newnumber{some} cases the authors of the benchmark have decreased the number of epochs in the buggy version from what was indicated in the source and in \newnumber{other} cases they have reduced the size of the training data. Both of these changes are likely to create a new additional root cause for low performance, interfering with the bug that is being reported. When it comes to the fixes applied in the benchmark, there were \newnumber{17} mismatches in this case. In \newnumber{around third} of those, the applied fixes were not proposed in any of the answers of SO posts or changes in GitHub commits, but were fully invented by the authors. The remaining fixes contain some changes picked from the suggestions but also introduce changes not suggested anywhere. 

For \textbf{DeepFD}, the buggy version corresponds to the source in \newnumber{23} cases,
and fixes match the ground truth fix of the source in \newnumber{21} out of \newnumber{36} cases. \newnumber{One common mismatch is that in some} faults in the dataset, the training data provided in the source is reduced by 20\% in the buggy version of the benchmark. Moreover, in the fixed version this reduction of the dataset is reverted. This leads to a mismatch with the ground truth fix, as the fix applied in the benchmark now includes an increase in training data. 

In the case of \textbf{Defects4ML}, the buggy versions provided in the benchmark generally show similarity to the source available (\newnumber{58} cases out of \newnumber{69}), while the fix matches the one/s provided in the source in \newnumber{79.7}\% of the analysed benchmark issues. In \textit{almost half cases} cases where the code is not the same as was reported in the source, the authors of the benchmark have artificially changed parameters like the original number of epochs, the batch size, activation or loss functions. In two bugs, the buggy version of the code already contained some of the fixes suggested in the source, while in another bug, the buggy and fixed version were swapped. 
In \newnumber{the majority of} cases where the fix does not correspond to the source, some additional or completely different fixes were applied to the buggy model when compared to the source.

When it comes to \textbf{TFBugs2018}, we found a relatively high match of the buggy model's code to the source (\newnumber{76.2}\%) and the highest match for the fixes (\newnumber{90.5}\%). The mismatch between the buggy code and its source range from manipulated hyperparameters such as learning rate and batch size to changes in loss function implementation.

Among all datasets, \textbf{SFData} has the lowest match rate of \newnumber{35.2\%} to the sources for the buggy version, while the match of the fixes is high at \newnumber{80.3\%}.
The mismatches for the fixes do not share any common characteristics, \newnumber{while the low rate for the match of the buggy versions is most likely due to the fact that the authors tried to reproduce tensor related faults without paying attention to many of the model parameters}. 

\begin{tcolorbox}[colback=white, colframe=black]
	\textbf{RQ1}: Our results show that, on average, bugs and fixes in the benchmarks match the sources \newnumber{52.8}\% and \newnumber{72.8}\% of the time. For two of the benchmarks, the match of buggy versions is around \newnumber{35-39}\%, indicating that nearly \newnumber{third} of the bugs were extracted inconsistently to their source.
\end{tcolorbox}

\subsection{RQ2 [Training Dataset]}

Column `\textit{TD m.}' in Table~\ref{tab:rq1_results} reports the number of cases in which the training data provided as the source of each bug matches the data used in the benchmark \newnumber{, and where this training data is available in the benchmark either as a separate file or through a code that downloads it.}. The correspondence to the original training dataset varies between \newnumber{52.4\%} and \newnumber{80.6}\% across the datasets. 
The main cause of mismatches is the lack of training dataset information in SO posts. SFData is most affected because it focuses on tensor shape faults, and therefore very often its faults are reported either using placeholders or randomly generated fake data.

The number of faults in each benchmark for which the used training data is real is reported in Column `\textit{TD real}' of Table~\ref{tab:rq1_results}. The instances of training data not being real stem from the previously discussed cases of the original training data not being mentioned or the SO posts using toy training data. For both DeepFD and DeepLocalize, the size of fake training datasets varies between \newnumber{12} and \newnumber{110,000} elements (they share a lot of the  SO posts). For TFBugs2018, the fake dataset can be as small as \newnumber{1} input and as big as \newnumber{88,041}. For Defects4ML, the smallest generated dataset has \newnumber{5} inputs, while the largest has \newnumber{60,000}. Lastly, the size of generated data for SFData varies between \newnumber{1} and \newnumber{412,457
} elements. 

\begin{tcolorbox}[colback=white, colframe=black]
	\textbf{RQ2}: The results reveal that training datasets used in the benchmarks are realistic only in \newnumber{42.6}\% of cases. Moreover, they correspond to the ones at the source of the bugs only in \newnumber{71.8}\% of cases, on average.
\end{tcolorbox}

\subsection{RQ3 [Fault Types]}

Table~\ref{tab:rq3_results_all} reports statistics on the composition of the bugs in the benchmarks, such as the maximum/average number of faults associated with each reported bug (i.e., the ``order'' of the bug, similarly to the order of a mutant in mutation testing), as well as their sum. 
Column \textit{`\#FT'} reports the number of unique fault types across the whole benchmark. The row \textit{`ALL'} contains information for the union of all benchmarks.  Note that the numbers indicated in brackets are for the subset of bugs that satisfy all the conditions for realism. 

\begin{table}[ht]
	\caption{Order of the bugs and fault types}
	\label{tab:rq3_results_all}
	\centering
	\begin{tabular}{c|c|c|c|c}
		\toprule
		\textbf{Dataset} & 
		\textbf{Max} &
		\textbf{Average} & 
		\textbf{Overall} &
		\textbf{\#FT}
		\\
		\midrule
		TFBugs2018 & 3 (2) & 1.2 (1.3)  & 25 (5) & 14 (4)\\
		DeepLocalize  & 6 (4) & 2.5 (2.3) & 93 (25) & 14 (9)\\
		DeepFD & 8 (4) & 3.2 (2.5) & 114 (5) & 15 (3) \\
		Defects4ML & 13 (7) & 2.1 (2.1) & 142 (60) & 37 (26) \\
		SFData & 3 (3) & 1.22 (1.3) & 173 (56) & 25 (17)\\      
		\midrule
		ALL & 13 (7) & 1.8 (1.7) & 547 (151) & 51 (39)\\  
		\bottomrule
	\end{tabular}    
\end{table}

The bugs in the benchmarks might have an order up to \newnumber{13}, with an average order of \newnumber{1.8} per bug. The benchmarks are also diverse when it comes to fault types, containing on average \newnumber{21} and all together \newnumber{52} different fault types. If we take into consideration only the faults that satisfy the condition for realism, the number of covered fault types goes down to \newnumber{39}. 

However, the level of representation of each fault type is quite disproportionate. For instance, out of \newnumber{547} fault components in \newnumber{305} bugs, the three most occurring fault types are wrong activation function (\newnumber{76}), wrong data pre-processing (\newnumber{69}) and missing layer (\newnumber{41}). Only \newnumber{9} fault types are represented in more than 20 fault instances and \newnumber{16} fault types have only one fault instance.  

\begin{tcolorbox}[colback=white, colframe=black]
	\textbf{RQ3}: The benchmarks include faults 
	with an average order of \newnumber{1.8}. The number of fault types in the benchmarks ranges from \newnumber{14} to \newnumber{36}, with all benchmarks containing a total of \newnumber{51} fault types.
\end{tcolorbox}

\subsection{RQ4 [Similarity to Artificial Faults]}

Table~\ref{tab:rq4_results} presents results on whether the bugs in the benchmarks can be simulated using existing mutation operators. Columns `\textit{RSB A.}' and `\textit{RSB R.}' report the \textit{ratio of bugs} that can be generated by applying the DeepCrime mutation tool to simulate each of its faults for the whole benchmark and for the subset of faults that satisfy the conditions for realism, respectively.  In contrast, columns `\textit{RSFC A.}' and `\textit{RSFC R.}'  provide a more fine-grained picture and report the \textit{ratio of fault components} that can be simulated. %

\begin{table}[ht]
	\caption{Results on similarity to artificial faults}
	\label{tab:rq4_results}
	\centering
	\begin{tabular}{c|c|c|c|c}
		\toprule
		\textbf{Dataset} & 
		\textbf{RSB A.} &
		\textbf{RSB R.} &
		\textbf{RSFC A.} &
		\textbf{RSFC R.} 
		\\
		\midrule
		TFBugs2018 & 0/21 (0\%) & 0/4 (0\%) & 0/25 (0\%) & 0/5 (0\%) \\
		DeepLocalize & 19/38 (50\%)& 6/11 (55\%)  & 64/93 (69\%)  & 18/25 (72\%) \\
		DeepFD & 19/36 (53\%) & 2/2 (100\%) & 89/114 (78\%)   & 5/5 (100\%)
		\\
		Defects4ML & 23/68 (34\%) & 8/29 (28\%) & 61/142 (43\%) & 26/60 (43\%)
		\\
		SFData & 1/142 (1\%) & 0/44 (0\%) & 4/173 (24\%) & 0/56 (0\%) \\      
		\midrule
		ALL & 62/305 (20\%) & 16/90 (18\%) & 218/547 (40\%) & 49/151 (32\%)\\  
		\bottomrule
	\end{tabular}
	
\end{table}

As the results show, only \newnumber{20}\% of the faults can be fully replicated by existing mutation operators. However, this number is highly affected by the fact that for the SFData and TFBugs2018 datasets almost none of the faults can be simulated. SFData focuses on tensor shape faults and there exist no mutation operators that are designed to mimic these faults. For TFBugs2018 the issue lies in the fact that its faults all use TensorFlow and the DeepCrime tool works only on the Keras framework. If we exclude SFData and TFBugs2018, \newnumber{43}\% of the bugs and \newnumber{61}\% of the fault components can be generated using existing mutation operators, with these numbers going up to \newnumber{53}\% and \newnumber{78}\% for the DeepFD dataset.

\begin{tcolorbox}[colback=white, colframe=black]
	\textbf{RQ4}: The results indicate that across analysed benchmarks \newnumber{20\%} of the faults can be simulated with the existing mutation operators.
\end{tcolorbox}

\subsection{RQ5 [Reproducibility]}
The number of bugs that have fulfilled our conditions on realism and which we therefore attempted to reproduce are reported in column \textit{`\#Bugs'} in Table~\ref{tab:rq5_results}. 
Our attempts were first challenged by the version problems. For most of the bugs, we were able to adapt the code to run on newer versions with minor adjustments. However, some bugs depended on outdated APIs and required a complete rewrite to be compatible with newer versions. The number of these issues is detailed in the `\textit{\#Version Problem}' column. The number of bugs that we could successfully reproduce is reported in column `\textit{\#Reproduced}'. Column `\textit{\#Stable}' reports the number of reproduced bugs that were stable across all 20 performed runs.

\begin{table}[ht]
	\caption{Reproducibility Results}
	\label{tab:rq5_results}
	\centering
	\begin{tabular}{c|c|c|c|c}
		\toprule
		\textbf{Dataset} & \textbf{\#Bugs} & \textbf{\#Version Problems} & \textbf{\#Reproduced}
		& \textbf{\#Stable}
		\\
		\midrule
		TFBugs2018 & 4 & 4 & 0 & N/A \\  
		DeepLocalize & 11 & 2  & 6 & 3\\
		DeepFD & 2 & 0 & 2 & 2 \\
		Defects4ML & 29 & 2 & 12 & 12 \\
		SFData & 44 &11 & 29 & 29 \\     
		\midrule
		ALL & 90 & 19 & 49 & 46\\  
		\bottomrule
	\end{tabular}  
\end{table}

As the results show, only \newnumber{54}\% of considered bugs are reproducible. TFBugs2018 and SFData are the only datasets that relied on very old Python and TensorFlow versions while not reporting any dependency requirements. Additionally, it was not possible to install old TensorFlow versions for 2 of Defects4ML bugs. These led to \newnumber{19} bugs left non-reproduced. Finally, from \newnumber{49} reproduced faults, newnumber{46} delivered results that are stable across 20 runs. 

A closer look into non-reproduced bugs reveals that for Defects4ML, running the provided buggy and fixed code led to errors in \newnumber{4} cases. Moreover, in 2 cases the repaired version had performance worse than the buggy and in 9 cases, the difference in performance between buggy and fixed was not statistically significant. For SFData, all cases of non-reproduced bugs were due to the buggy version not producing any tensor shape error. 

\begin{tcolorbox}[colback=white, colframe=black]
	\textbf{RQ5}: Out of \newnumber{90} bugs, we were able to successfully reproduce \newnumber{49} and the behaviour of \newnumber{46} of them was stable across all the performed runs. 
\end{tcolorbox}

\section{Threats to Validity}

\textbf{Construct.} The main threat to the construct validity is our choice of characteristics to consider a DL fault in a benchmark as ``real''. To minimise this threat we use straightforward and unambiguous criteria such as correspondence to source for training data and buggy/fixed versions. We also provide a precise definition for what we consider a realistic training data. 

\noindent \textbf{Internal.} One possible threat to the internal validity is that our literature analysis may not include all relevant works. To ensure that we covered the list of works referring to reproducible real DL faults, we employed a comprehensive search string and performed both forward and backward snowballing.

Another possible threat is the possibility of human error when performing the manual analysis. To minimise this risk we have conducted a pilot with two of the authors to align their understanding of the task. We have also employed automated tools for comparison of the files. Lastly, the authors took note of any non-trivial issues and ensured the decision-making for those cases was performed via consensus. 
\section{Discussion}

\subsection{Towards More Realistic Benchmarks}

\subsubsection{Correspondence to Source}

Our analysis across five established deep learning fault benchmarks reveals a significant gap in the correspondence between reported faults and their actual buggy and fixed source versions. On average, only 41.6\% of the faults can be confidently matched to their original buggy and fixed code. Notably, in one of the benchmarks, this number drops as low as 16.7\%. When we apply a stricter definition of realism, requiring that the training data also corresponds to the source it was derived from and reflects realistic usage, the average proportion of qualifying faults drops sharply to just 15.4\%. These findings indicate that current DL fault benchmarks often do not ensure a reliable match between reported faults and their true sources, while this correspondence is a core element of realism in traditional software engineering fault benchmarks. 

We believe our findings offer strong support and clear guidance for future research efforts that aim to collect new benchmarks or use existing ones for evaluation purposes. In particular, studies that construct new benchmarks should provide a detailed description of their experimental procedures and explicitly define what they consider to be a real fault. Our four proposed realism criteria can serve as a valuable starting point for such analysis. Researchers can use these criteria as a checklist to guide the design of their benchmarks, or as a framework for transparency, specifying any deviations from the criteria and explaining why those deviations are acceptable in their context and do not compromise realism.

For studies that rely on existing benchmarks that we have analysed, the results of our labelling that is made publicly available can inform more rigorous fault selection. Researchers may choose to use only the faults that satisfy all four realism criteria or apply filtering based on specific criteria they consider essential for their evaluation. Moreover, our analysis can also serve educational purposes. In particular, users can focus on the faults we were able to reproduce to gain insights into fault types, typical fixes, and repair strategies, providing a more detailed and instructive view of deep learning program faults and their resolution.

\begin{tcolorbox}[colback=white, colframe=black]
	\textbf{Insight 1:} Our results highlight a widespread issue in DL fault benchmarks: many reported faults cannot be reliably traced back to their actual buggy and fixed code versions. To improve the quality and transparency of future evaluations, we recommend that researchers clearly define what they consider a real fault and use structured criteria to guide or document their benchmark construction. Our publicly available dataset and proposed realism criteria offer practical tools to support this effort.
\end{tcolorbox}

\subsubsection{Representativeness}
During our analysis, along with the data required to answer our research questions, when available, we also collected data on what type of task (e.g. classification or regression) the DL model associated with the bug is focused on and what DL framework it uses. As the results in Table~\ref{tab:disc_repr} show, in \newnumber{71.5}\% of the cases the task was classification (Column 'C'), while in 14.75\% of the faults the task being solved was regression. During our analysis, in 7 cases the task being solved was unclear, and in another 7 faults no task was being solved in the code. In 28 issues, we observed a variety of task types, for example, Image Manipulation' (4), Image De-noising' (3), Dimensionality Reduction' (2), and `Text Generation' (1). The full list is available in the replication package~\cite{replpackage}. When it comes to DL frameworks, 2 out of 5 benchmarks focus specifically on Keras, one on TensorFlow (TF), while the remaining two use both Keras and TF. This translates into 71\% of the faults implemented with Keras and 29\% with TF. 

\begin{table}[ht]
	\caption{Results on performed task types and used frameworks }
	\label{tab:disc_repr}
	\centering
		\begin{tabular}{c|c|c|c|c|c}
			\toprule
			\textbf{Dataset} & \multicolumn{3}{c}{\textbf{Task Type}} & \multicolumn{2}{c}{\textbf{Framework}} \\
			& \textbf{C} &
			\textbf{R} &
			\textbf{O} &
			\textbf{Keras} &
			\textbf{TF} 
			\\
			\midrule
			TFBugs2018 & 15 (79\%)& 4 (21\%)& 0 (0\%)& 0 (0\%)& 21 (100\%)\\
			DeepLocalize & 28 (74\%)& 9 (24\%)& 1 (2\%)& 38 (100\%)& 0 (0\%)\\
			DeepFD & 26 (74\%)& 9 (26\%)& 0 (0\%)& 36 (100\%)& 0 (0\%)\\
			Defects4ML & 53 (78\%)& 9 (13\%)& 6 (9\%)& 57 (84\%)& 11 (16\%)\\
			SFData & 96 (73\%)& 14 (11\%)& 21 (16\%)& 85 (60\%)& 57 (40\%)\\      
			\midrule
			Overall & 218 (75\%)& 45 (15\%)&  28 (10\%)& 216 (71\%)& 89 (29\%)\\  
			\bottomrule
		\end{tabular}
	
\end{table}

We also checked the bugs that use realistic datasets and collected the names of such datasets. Overall, \newnumber{90} different bugs use \newnumber{21} different well-estanblished and publicly available datasets. Of them, \newnumber{52} of the bugs use MNIST, \newnumber{11} use CIFAR-10, \newnumber{8} use a dataset for XOR boolean function, 5 use Iris dataset and \newnumber{3} use IMDB. Out of the remaining \newnumber{16} datasets, \newnumber{7} are used in 2 bugs and \newnumber{10} are used in only one bug.

\begin{tcolorbox}[colback=white, colframe=black]
	\textbf{Insight 2:} We observe that most existing benchmarks are heavily centered on the Keras framework and classification tasks involving simple datasets like MNIST and CIFAR-10. This narrow focus limits representativeness, highlighting the need for benchmarks that include more diverse task types, such as regression, and incorporate more complex and varied datasets.
\end{tcolorbox}

\subsubsection{Sources of Bugs}

84\% of the faults we analysed originate from SO and only 16\% come from GitHub. It should be noted that using SO as a source of bugs is something specific to DL research, as for traditional software systems, bugs are commonly mined from GitHub. This is mostly due to the fact that GitHub repositories for DL models very often do not include intermediate points in the DL model development process, but are committed once a well-performing model has been trained. 
While the usage of SO is understandable in terms of convenience and providing information on such intermediate problems, it comes with a lot of drawbacks. As SO is a discussion forum,  posts very often do not contain the full code and information on the used training dataset. This creates a need for the benchmark authors to fill the gaps. As the symptom for DL models is usually also indicated in very imprecise terms (such as low accuracy, slow training process, etc.) it is hard to ensure that the changes introduced to fill such gaps match the root causes of the reported symptoms. As there is a low rate of match between the buggy version and the initial source code in SO, our assumption that the information in the posts is very often not sufficient to reproduce the reported problem. Hence the authors sometimes used SO posts as an inspiration of an example code and then introduced some changes that reproduce to some degree the reported symptoms. The same applies to the performed fixes that also have a low match rate to the answers provided in SO, most likely due to the vagueness in the answers and their failure to really fix the fault. While faults generated in such a manner might indeed provide pairs of buggy and fixed models, this approach is not very different from (manual) mutation testing and such benchmarks create a false promise of real bugs.

We believe the usage of SO for the purpose of extracting real bugs can only be reliably performed if strict criteria are applied to the posts: relevant posts must report the whole code and indicate the training dataset used, as well as contain an accepted answer proposing a precise fix solution. However, we understand that the number of such posts is limited and that their identification requires significant manual effort.

\begin{tcolorbox}[colback=white, colframe=black]
	\textbf{Insight 3:} Most DL faults in benchmarks come from StackOverflow (84\%). While SO posts offer convenience, they often lack complete code and dataset details, making faithful reproduction difficult. As a result, benchmark authors must fill gaps and approximate fixes, risking the creation of artificial faults similar to manual mutations. To improve reliability, stricter criteria should be applied to select SO posts that provide full code, dataset information, and accepted precise fixes.
\end{tcolorbox}

Another problem with extracting DL faults is that GitHub and SO are not the sources that can cover all possible types of faults. This is demonstrated by the taxonomy of real faults in DL systems by Humbatova et al.~\cite{Humbatova:2020}, which has used these two sources to analyse faults, but discovered many additional fault types as a result of interviewing developers. Therefore, collecting a diverse set of real faults requires a close inspection of the DL development process by developers. One approach to cover types of faults not represented in GitHub and SO might be conducting developer studies where intermediate points in DL development are recorded. Such studies are very expensive and would require careful planning, but would provide an in-depth view on the evolution of DL models and would lead to the creation of a dataset of faults that have indeed taken place during the DL development process.

\begin{tcolorbox}[colback=white, colframe=black]
	\textbf{Insight 4:} GitHub and StackOverflow, while commonly used, are insufficient to capture the full spectrum of real faults in deep learning systems. To build more comprehensive and representative benchmarks, future work should explore complementary methods such as developer studies that document faults throughout the development lifecycle. 
\end{tcolorbox}

\subsubsection{Benchmark Independence} 4 out of the 5 benchmarks that we analysed have been generated specifically to evaluate the approach proposed by the authors of the benchmark. In the absence of any dataset of DL faults, this is an understandable choice. Moreover, the authors provide a systematic procedure on how  SO and GitHub issues have been mined. However, since for many faults, both the buggy and the fixed versions have changes not indicated in the respective source, there is a risk of unconscious bias towards introducing faults and fixes that would highlight the strengths of the proposed approach. For this reason, it is important that independent benchmarks such as Defects4ML are created and maintained, to allow for an evaluation on a dataset not generated by the same authors who are proposing and evaluating a novel approach.

\begin{tcolorbox}[colback=white, colframe=black]
	\textbf{Insight 5:} Most DL fault benchmarks are created by the authors of the evaluated DL testing techniques, which risks biasing faults and fixes to favor those methods. Independent benchmarks like Defects4ML are needed to provide unbiased evaluation datasets developed separately from the approaches being tested.
\end{tcolorbox}

\subsubsection{Benchmark Maintainability} Our attempts to reproduce the bugs in these benchmarks show problems with the maintainability of the bugs, as they rely on old versions of Python and the DL frameworks and often can not be run with more recent versions and hardware. One effective strategy to address this problem is containerisation using platforms like Docker. This would allow for preserving the conditions under which a bug was originally identified and ensure that legacy software in these bugs remains functional despite updates to the underlying frameworks.

\begin{tcolorbox}[colback=white, colframe=black]
	\textbf{Insight 6:} The analysed benchmarks often suffer from poor maintainability due to outdated dependencies. Containerisation (e.g., via Docker) is a practical solution to preserve reproducibility and long-term usability.
\end{tcolorbox}

\subsection{Mutants vs. Real Faults}
Our results have important implications for mutation testing tools of DL systems. As mentioned before, the existing tools are capable of applying only first-order mutants, while our analysis of real faults indicates that \newnumber{38}\% of faults have a higher order. Therefore, adding the feature of generating higher-order mutants can help the existing tools generate faults that are more similar to the real ones. However, applying all possible combinations of single-order mutants is expensive. Our analysis of fault types in each bug can be used as a data source to study the most frequent combinations that then could be integrated in the higher-order mutation generation process. 

As discussed before, the representativeness of the benchmarks in terms of subjects/datasets is quite limited and the distribution of faults is heavily skewed towards specific fault types. While improving the quality of benchmarks with respect to these weaknesses is an important and worthwhile direction to take, using mutations in the evaluations is another approach that can be applied to allow a more thorough evaluation. Mutation tools can be applied to any DL system and by applying different types of mutation operators various fault types can be investigated. To the best of our knowledge, the work by Kim et al.~\cite{kim2023repairing} is the only one that performs an evaluation on both mutants and a subset of real faults from DeepFD's dataset. Their results show that program repair tools perform better on real faults than mutants, as mutations are applied to more complex DL models, trained on larger datasets. 

\begin{tcolorbox}[colback=white, colframe=black]
	\textbf{Insight 7:} Existing DL mutation testing tools are limited to first-order mutants, while many real faults are higher-order. Incorporating common higher-order fault patterns and applying mutations across diverse DL systems can improve the representativeness and usefulness of mutation-based evaluations.
\end{tcolorbox}

\section{Related Work}

Humbatova et al.~\cite{Humbatova:2020} constructed a comprehensive taxonomy of DL faults by interviewing 20 developers and manually analysing 1,059 artefacts from GitHub and SO. Zhang et al.~\cite{zhang2018issta} focused on bugs in TensorFlow and examined 175 bugs collected from SO and Github. While our work is not focused on taxonomising real DL faults, we also analysed the fault types present in previous studies that constructed fault datasets, to understand the fault type distribution.

Cao et al.~\cite{cao2025study} performed an empirical study to evaluate the capabilities of LLMs in DL program repair. Their evaluation relies on the existing DeepFD benchmark and focuses primarily on prompt design and repair performance, without analysing the realism of the faults contained in the benchmark.

There have been recent efforts to address the reproducibility issues in DL. Kim et al.~\cite{kim2023repairing} attempted to reproduce faults in the DeepFD benchmark and identified that only 9 out of 58 faults were reproducible. Their analysis attributed this to two factors: cases where the fixed version did not improve the performance over the buggy version, and instances where the related issue did not expose the fault in the buggy version. Shah et al.~\cite{shah2025towards} conducted an empirical study on the reproducibility of DL bugs and identified ten key edit actions, such as `Adding logging statements' and `Initialising hyperparameters', which facilitated the reproduction of these bugs. In contrast, our work presents the first comprehensive analysis of the realism and reproducibility of DL faults, taking into account a broader range of factors. Furthermore, we apply this analysis beyond DeepFD, considering in total 490 bugs across five DL fault benchmarks introduced in previous works.
\section{Conclusion}

We have conducted an empirical study on the realism and reproducibility of existing DL fault benchmarks.
We manually analysed 490 bugs from 5 frameworks and identified that only 47 of them satisfy our conditions on realism. Our attempts at reproducing the bugs were successful in 54\% of cases. 

To improve the existing benchmarks or produce new ones, researchers should apply stricter inclusion criteria when mining DL bugs and should go beyond repository mining, e.g., by conducting controlled experiments or observational studies to collect bugs from the field; they should invest on the creation of independent benchmarks that can be reused by the authors of novel DL testing approaches, to avoid the authors' bias; they should aim at covering all different fault types; they should consider DL mutation tools as a complementary assessment benchmark for novel approaches.
Moreover, our discussion on the correspondence of benchmark bugs to mutants provides clear points on how mutation tools can be enhanced to produce artificial faults more similar to real ones. 

\section{Data Availability} \label{sec:data}
The experimental data, and the evaluation results that support the findings of this study are available at the following link:

https://github.com/dlfaults/HowRealAreThey

\bibliographystyle{IEEEtran}
\bibliography{ref}

\end{document}